\documentclass[aps,showpacs,prl,twocolumn]{revtex4}
\usepackage{epsfig,amssymb,amscd,graphicx}
\usepackage{subfigure}
\usepackage{psfrag}

\newcommand{\nin}{\noindent}
\newcommand{\non}{\nonumber}

\newcommand{\bea}{\begin{eqnarray}}  
\newcommand{\eea}{\end{eqnarray}} 
\newcommand{\be}{\begin{equation}}  
\newcommand{\ee}{\end{equation}}  
\newcommand{\braket}[2]{\langle #1|#2\rangle}
\newcommand{\ket}[1]{ | \, #1  \rangle} 
\newcommand{\bra}[1]{ \langle #1 \,  |}

\begin{document}

\title{Eigensolutions of the kicked Harper model}
\author{G. A. Kells}
\affiliation{Department of Mathematical Physics, National University of Ireland, Maynooth,  Ireland.}

\begin{abstract}
The time-evolution operator for the kicked Harper model is reduced to block matrix form when the effective Planck's constant $\hbar = 2\pi \frac{M}{N}$ and $M$ and $N$ are integers. Each block matrix is spanned by an orthonormal set of $N$ $kq$ (quasi-position/quasi-momentum) functions. This implies that the system's eigenfunctions or stationary states are necessarily discrete and periodic. The reduction allows, for the first time, an examination of the 2-dimensional structure of the system's quasi-energy spectrum and the study of, with unprecedented accuracy, the system's stationary states.  \end{abstract}


\date{\today}
\maketitle

\section{Introduction}

The Harper model was first introduced in 1955 to approximate the dynamics of electrons confined to a 2-dimensional lattice while under the influence of a perpendicular magnetic field \cite{har55}. The model has since been generalised with the addition of a periodic kicking term so that it is classically chaotic under certain conditions \cite{zas86a,afa90}.  The dynamics of this {\em kicked} Harper model can be related to those of the kicked harmonic oscillator when a ratio of $1/4$ exists between the kicking and oscillation frequencies. 

The kicked harmonic oscillator was originally proposed as a 2-dimensional model of charges moving in a homogeneous static magnetic field while under the influence of an orthogonal time-dependent electric field \cite{zas86b,che88,lic89,zas91,dan95a}. The model is fundamentally different from other widely studied kicked systems because the natural frequency of the unperturbed system does not depend on energy. It therefore cannot be described using the KAM theorem \cite{kol54}. The quantized system has been proposed as a model for electronic transport in semiconductor super-lattices \cite{fro01} and for atom optic modeling in ion-traps \cite{gar97}. 

The quantum kicked harmonic oscillator and the quantum kicked Harper model, in the same way as the quantum kicked rotator \cite{lev03}, may be simulated efficiently on a quantum information processor \cite{kel04,lev04}, using the Quantum Fourier Transform \cite{sho97} and Quantum Fractional Fourier Transform \cite{kla03}.

\begin{figure}[htbp]  
       \includegraphics[width=.3\textwidth,height=0.3\textwidth]{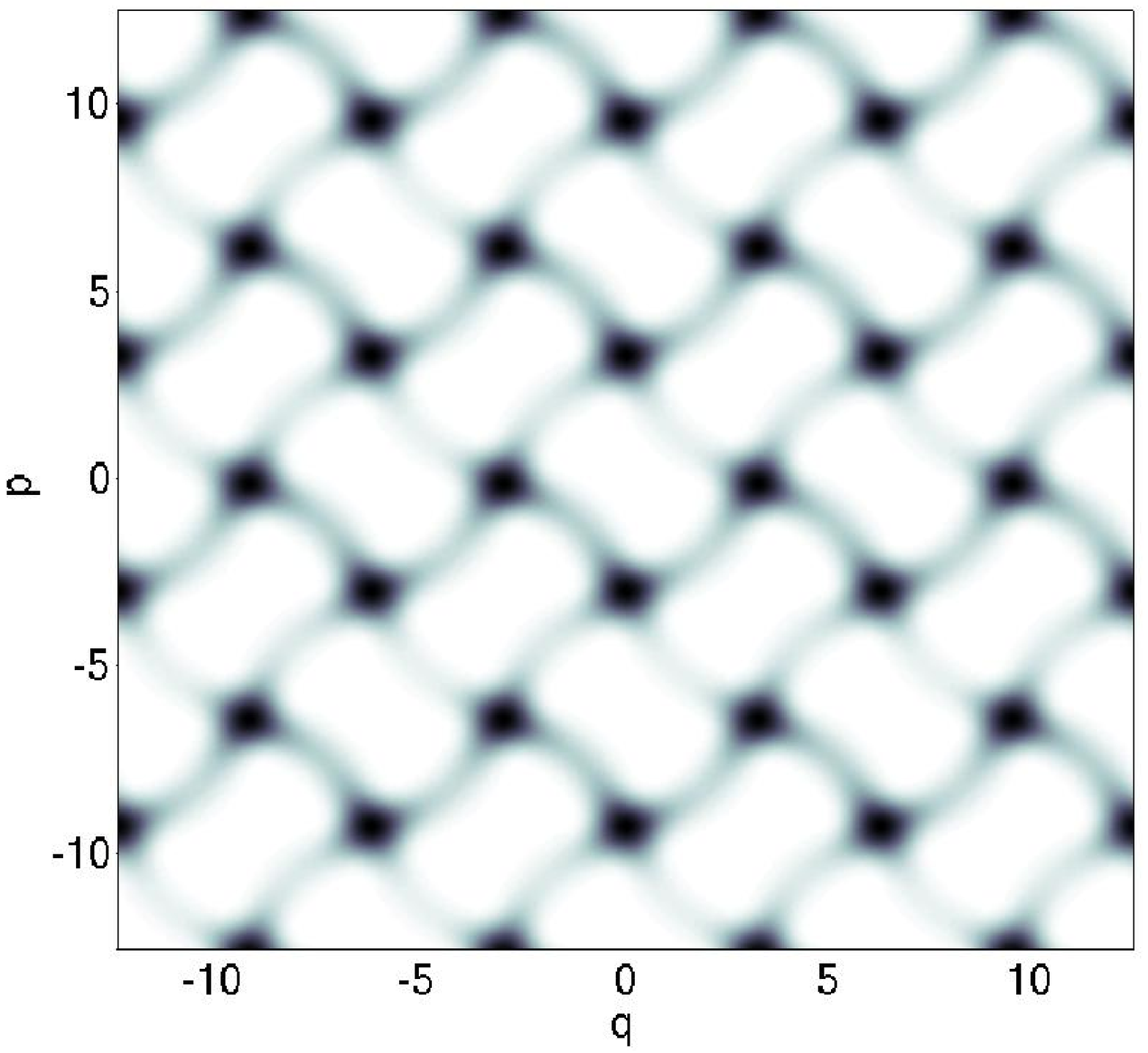}
       \caption{A section of the Husimi distribution of a quantum stationary state of the kicked Harper model when $K=L=1.5$ and $\hbar =2 \pi /20$.  The distribution has a structure similar to the Poincar\'{e} surface of section illustrated in Fig. \ref{fig:eigfunc_webcl}. } 
       \label{fig:eigfunc_webq}
\end{figure}

In this article we show how the evolution operator for Kicked Harper model may be reduced to block matrix form in the {\em kq} representation. This is then used to examine the 2-dimensional structure of the systems quasi-energy bands as well as to construct the eigenfunctions of the evolution operator in the position and momentum representations. We also compare the structure of selected eigenfunctions with the classical phase-space using the Husimi quasi-probability distribution, see Fig. \ref{fig:eigfunc_webq}.

\section{The Kicked Harper Model}

The generalised kicked Harper Hamiltonian is written as 

\be
H= W(p)+V(q)\sum_{j=-\infty}^{\infty}\delta(t-j).
\label{eq:Harper}
\ee

\nin 
It is usually assumed that $W(p) =L \cos p$ and $V(q)=K\cos q$ although our analysis will apply to all functions that are $2\pi$ periodic. 

\begin{figure}[htbp]    
       \centering
       \includegraphics[width=.30\textwidth,height=0.3\textwidth]{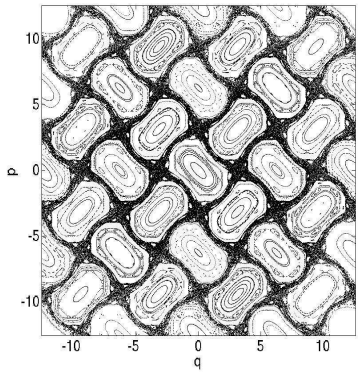}
       \caption{Poincar\'{e} surface of section for the kicked Harper model when $K=L=1.5$. }
       \label{fig:eigfunc_webcl} 
\end{figure}

The classical model displays some unusual phase space properties. In particular is the ordering of the Poincar\'{e} surface of section into a lattice of stable elliptical cells when the perturbation is applied. Each cell is separated by a mesh of unstable trajectories which are collectively called the {\em stochastic} web, see Fig. \ref{fig:eigfunc_webcl}. The web extends to all areas of the phase plane and grows in thickness as the perturbation strengths are increased \cite{afa90}.

Another interesting feature of the model is the existence, under certain resonant conditions, of accelerator modes or islands. These modes allow for stable ballistic motion throughout the phase space and are the main reason behind what is called anomalous diffusion or anomalous energy growth \cite{afa90,dal95, zas97}.  

The quantum mechanical study of this problem has been dealt with in many guises and by many authors \cite{leb90,ber91,art92,gua93,art94,bor95a,bor95b,dan95b,dan95c,fra97,ket99,sat05}. The delta function in (\ref{eq:Harper}) means that the quantum mechanical time-evolution or Floquet operator may be factored as

\be
U(q,p) = P \times Q = \exp \left(-\frac{i}{\hbar}W(p) \right)  \exp \left(-\frac{i}{\hbar} V(q) \right) ,  
\label{eq:U_KHM}
\ee

\nin
where $p$ and $q$ are the momentum and position operators whose eigenvalue equations are $p\ket{p'} = p'\ket{p'}$ and $q\ket{q'} = q'\ket{q'}$ respectively. The Fourier transform operation, 

\be
\mathcal{F} f(q') = \frac{1}{\sqrt{2 \pi\hbar}} \int \exp \left(-\frac{i}{\hbar} q' p' \right) f(q') dq', 
\ee

\nin
provides the means to write (\ref{eq:U_KHM}) in either position or momentum representations. If it is understood that $\mathcal{F}$ acts on everything to the right we may say that  $P(p)= \mathcal{F}^\dag P(q) \mathcal{F}$ where $P$ is a function of the operators $p$ and $q$. We can therefore write the evolution operator $U$ as

\be
U(q) = \mathcal{F}^\dag \exp \left(-\frac{i}{\hbar}W(q) \right) \mathcal{F} \exp \left(-\frac{i}{\hbar} V(q) \right) . 
\label{eq:U} 
\ee 

The system's link with the Kicked Harmonic Oscillator can be seen from the fact that the the Floquet operator of that system can be written as

\be
U_{KHM}(q,\theta) = \exp \left(-i \frac{\theta}{2} \right)  \mathcal{F}_{-\theta} \exp\left(- \frac{i}{\hbar} V(q) \right), 
\ee

\nin
where $\mathcal{F}_{-\theta}$ represents what is called the fractional Fourier transform and is, up to a phase, the evolution operator of the simple harmonic oscillator in the position representation \cite{nam80,kel04} . The ordinary Fourier transform is defined as $\mathcal{F} \equiv \mathcal{F}_{-\frac{\pi}{2}}$ and time-evolves a state forward $1/4$ of one full oscillation.

\vspace{5mm}

\section{The Floquet operator in the $kq$ representation}

The operators $U$ and $U_{KHM}^4(q,\frac{\pi}{2})$ can both be shown to be simultaneously translationally invariant in position and momentum if $\hbar = 2\pi M/N$ \cite{bor95b,dan95c}. This suggests that the orthonormal $kq$ (quasi-position/quasi-momentum) functions of Zak \cite{zak67}, which is also translationally invariant in position and momentum, would be a convenient representation to work in.

To define this representation we first introduce the Dirac delta comb or train as

\be
\braket{q'}{\text{III}_l} \equiv \text{III}_l(q') \equiv \sum_{n=-\infty}^{\infty} \delta ( q' - n l).  
\ee 

\nin
Then, in a similar way to Dana \cite{dan95c}, we consider the (slightly modified) {\em kq} functions, which we define as

\be
\label{eq:pos_rep}
\braket{q'}{r,s,l} \equiv C_q\exp \left(\frac{is}{\hbar}  (q'-\frac{r}{2}) \right)\text{III}_l(q'-r)
\ee 

\nin
with $C_q= \sqrt{\frac{l}{\sqrt{2\pi \hbar}}}$. In our notation the $k$ of Zak is replaced by $r$ and $q$ is replaced by $s$ so we may use $q$ and $q'$ as the position operator and variable without confusion. It has been shown that these functions form an orthonormal set with the labels $0 \le r < l$, $ 0 \le s <a$  and $a=2\pi\hbar/l$ \cite{zak67}.

In the momentum representation the states $\ket{r,s,l}$ may be written as

\bea
\label{eq:mom_rep}
\braket{p'}{r,s,l} &\equiv& \mathcal {F} \braket{q'}{r,s,l} ,\\ \non &=& \frac{1}{\sqrt{2\pi\hbar}} \int \exp \left(-\frac{i}{\hbar} q' p' \right) \braket{q'}{r,s,l} dq' ,  \\ \non  &=& C_p \exp \left(\frac{-ir}{\hbar}  (p'-\frac{s}{2}) \right)\text{III}_a(p'-s), 
\eea 

\nin where $C_p= \sqrt{\frac{a}{\sqrt{2\pi \hbar}}}$ \cite{poisson}. It should be noted that the modification to Zak's definition is solely to preserve the symmetry between momentum and position representations.

The procedure for reducing expression (\ref{eq:U_KHM}) to block matrix form is a technical exercise in labeling and reordering and therefore given in the appendix. There it is shown that for block reduction to be possible one must choose the inversely related $l$ and $a$ so that they are both rational multiples of the periodicity of $W$ and $V$. For $\hbar = 2\pi M/N$ one may do this by either setting $(l,a) = (2 \pi M/N_s, 2 \pi/N_r)$ or $(l,a) = (2 \pi /N_s, 2 \pi M /N_r)$ with the added constraint that $N_r N_s = N$. The resulting block matrices are of dimension $N \times N$ and are spanned by the kets 

\be
\ket{r+j\hbar,s+k\hbar,l}
\label{eq:span}
\ee

\nin  where $j \in {0,1,...,N_r-1}$ and  $k \in {0,1,...,N_s-1}$. Importantly, this set of kets defines the subspaces on which the system dynamics take place.

As an example suppose $\hbar=\pi/2$. One can make the choices $(l,a)= (\pi,\pi),(2\pi,\pi/2)$ or $(\pi/2,2\pi)$ among others. The first choice $(l,a)= (\pi,\pi)$ means the spanning subspaces are

\bea
&&\ket{r,s,\pi} \non \\
&&\ket{r,s+ \frac{\pi}{2},\pi} \non \\
&&\ket{r+ \frac{\pi}{2},s,\pi} \non \\
&&\ket{r+\frac{\pi}{2},s+ \frac{\pi}{2},\pi}.
\eea

\nin
for $0 \le r < \hbar$ and $ 0 \le s <\hbar$. The second and third choices mean the subspaces are

\bea
&&\ket{r,s,2 \pi} \non \\
&&\ket{r+ \frac{\pi}{2} ,s,2 \pi} \non \\
&&\ket{r+ \pi,s,2\pi} \non \\
&&\ket{r+\frac{3\pi}{2},s,2\pi},
\eea

\nin and  

\bea
&&\ket{r,s,\frac{\pi}{2}} \non \\
&&\ket{r,s+\frac{\pi}{2},\frac{\pi}{2}} \non \\
&&\ket{r,s+\pi,\frac{\pi}{2}} \non \\
&&\ket{r,s+\frac{3\pi}{2},\frac{\pi}{2}},
\eea

\nin
respectively, where again $0 \le r < \hbar$ and $ 0 \le s <\hbar$.  It can be shown that these three subspaces are actually one and the same, see \cite{rev05}. In the appendix, the elements of each block matrix are calculated and written in a concise form using the generalised Fourier matrices, see equations (\ref{eq:xPxQ}) and (\ref{eq:PxQx}). As each block matrix is spanned by the $kq$ states of (\ref{eq:span}), solutions to the eigenvalue equation  

\be
U\ket{\psi} = e^{i\omega} \ket{\psi},
\ee

\nin
must be of the form

\be
\ket{\psi} = \sum_{j =0}^{N_r -1} \sum_{k=0}^{N_s-1} \psi_{j,k} \ket{r+j\hbar,s+k\hbar,l},
\label{eq:eig}
\ee

\nin
where the $\psi_{j,k}$ are the $N$ elements of one eigenvector of either (\ref{eq:xPxQ}) or (\ref{eq:PxQx}).  This implies that generic eigenfunctions of $U$, with $\hbar=2 \pi M/N$,  are translationally invariant in position and momentum. This is true regardless of the values of $K$ and $L$.  

This result has important consequences for the interpretation of the systems dynamical behaviour, in particular for the behaviour of initially localised states over long time scales. To measure the spread of a wavefunction over time one often examines the harmonic oscillator energy function $E=\frac{1}{2}(q^2+p^2)$. One can classify the type of dynamics observed in terms of the growth rate of the energy expectation value  $\bra{\phi_n}E\ket{\phi_n}$ where $\ket{\phi_n} = U^n \ket{\phi_0}$. 

 A significant amount of effort has been made in recent decades to try and understand these growth rates in terms of system's quasi-energy distribution and the type of stationary states observed. The result above is important because it says that all eigenstates (at least when $\hbar=2 \pi M/N$) are extended and therefore suggests that qualitative differences in observed dynamical behaviour should depend on the quasi-energy spectrum alone. However, this is not the full story. While the result deprives us the use of the eigenfunctions to explain the dynamics it provides us with another means, namely the size of the dynamical subspaces $N$.    

 Analytical and numerical arguments by Borgovoni {\em et al.} \cite{bor95b} have shown that if $\hbar = 2\pi \rho$, where $\rho$ is some irrational number, then the energy growth rate after a long time is {\em linear}. However they also show if $\hbar = 2 \pi M /N $, the energy growth rate eventually becomes {\em quadratic} in time. This quadratic growth is sometimes known as quantum {\em resonance} and has been confirmed by a number of different numerical techniques \cite{eng03,kel05}. 

Although we cannot use the $kq$ functions to examine the non-resonant situation $\hbar = 2\pi \rho$ (there being no way to choose $(l,a)$ so that they both are rational multiples of $2\pi$),  we can always say that there exists a rational approximate $M/N$ that is {\em arbitrarily} close to $\rho$ and for which {\em kq} block reduction can be achieved. With this idea in mind it may be possible to relate the {\em time} before the onset of quadratic energy growth to the {\em sizes} $N$ of the extended but {\em finite} subspaces. The constant linear energy growth rate observed for $\hbar = 2 \pi \rho$ could then be explained as a direct consequence of the motion taking place on an extended but {\em infinite} space. 

The form of the Floquet eigenfunctions, (\ref{eq:eig}), strongly suggests that the dynamical localisation observed in the kicked rotator \cite{cas79} does not occur in this model. Dynamical localisation is when the quantum expectation value $\bra{\phi_n}E\ket{\phi_n}$ saturates around some constant value while the energy of a classical ensemble, with the same initial marginals, continues to grow \cite{haa01}. The most widely cited explanation of this process is given by Gremple {\em et al.} \cite{gre82}. Here the phenomena is linked to Anderson localisation \cite{and58} by way of a mapping to a tight-binding system. However, a key point in this explanation is that the Floquet eigenfunctions are exponentially localised. Clearly the eigenfunctions of the kicked Harper model do not display this property. 

Indeed, the form of the eigenfunctions (\ref{eq:eig}) suggests that, for all non-zero values of $K$ and $L$, indefinite localisation is unlikely. Note that after each operation of $U$, and because of differing quasi-energies, the interference between Floquet eigenfunctions that initially allowed the creation of a localised state, will gradually disappear. The fact that the dynamics take place on extended subspaces means that the wavefunction must eventually begin to spread. This should happen even if the initial state is placed inside a stable {\em classical} phase-space structure. 

The only way this dispersion or diffusion may be prevented is if the quasi-energies of the states that make up the superposition are all the same or only differ from each other by some rational multiple of $2\pi$. It is difficult to envisage a situation where this could be arranged 

\section{Application}

In this section some analytical and numerical examples of eigensolutions for specific $K,L$ and $\hbar$ are given. These are primarily included to instruct and to visually aid understanding. However they are also to demonstrate the consistency and validity of the results of the previous section and of the analysis given in the appendix.

\begin{figure}
     \subfigure[$\omega_\pm(r,s)$ with $L=\pi,K=\pi$]{
           \label{fig:omeag_pia} 
          \includegraphics[width=.3\textwidth,height=0.23\textwidth]{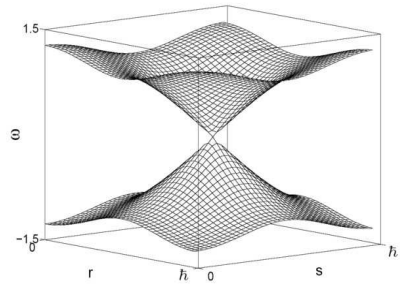}}
     \subfigure[$\omega_{+}(r,s)$ with $L=\pi,K=\pi$ ]{
           \label{fig:omega_pib}
          \includegraphics[width=.3\textwidth,height=0.23\textwidth]{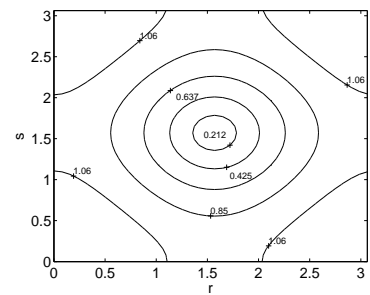}}
     \subfigure[$\omega_\pm(r,s)$ with $K=10,L=11$]{
           \label{fig:omega_pic}
          \includegraphics[width=.3\textwidth,height=0.23\textwidth]{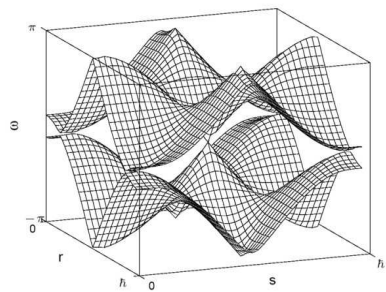}}
      \subfigure[$\omega_{+}(r,s)$ with $K=10,L=11$]{
           \label{fig:omega_pid}
          \includegraphics[width=.3\textwidth,height=0.23\textwidth]{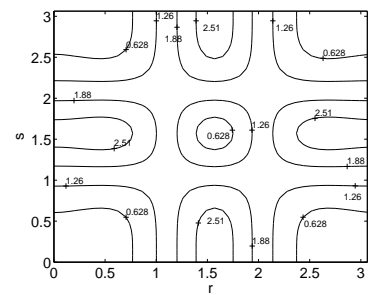}}
     \caption{$\omega_\pm(r,s)$ ( see eq. (\ref{eq:eig_hbar_pi}))  for $\hbar =\pi$ and various values of $K$ and $L$. }
     \label{fig:omega_pi}
\end{figure}

\subsection{Explicit calculation with $\hbar= \pi$}

One can apply (\ref{eq:xPxQ}) or (\ref{eq:PxQx}) to the problem with $\hbar=\pi$ ($M=1$ and $N=2$). This has already been done in \cite{sat05}, however we may now give the results in terms of the two parameters $r,s$ instead of just one. Using (\ref{eq:xPxQ}) it can be seen that

\bea
\label{eq:Ursl}
U(r,s) &=&   X(1,2) \mathcal{P}(r,s) X(1,2) \mathcal{Q}(r,s)  \\&=& \left( \begin{array}{cc} \exp(- i\bar{L})\cos( \bar{K})  &-i\exp( - i\bar{L}) \sin(\bar{K})\theta  \\ -i \exp( i\bar{L}) \sin(\bar{K}) \theta^*  &\exp( i\bar{L})\cos( \bar{K})  \end{array} \right) \non
\eea


\nin
where $\bar{L}=\frac{L}{\hbar} \cos s$, $\bar{K}=\frac{K}{\hbar} \cos r$ and $\theta=\exp(i r/2)$. This calculation takes place on the subspace

\be
\non 
\ket{r,s,l}=\left( \begin{array}{c} 1 \\ 0  \end{array} \right) ,\quad  
\ket{r,s+\hbar,l} =\left( \begin{array}{c} 0\\ 1  \end{array} \right)\;\; ,
\ee 

\nin but one could have also defined the subspaces  

\be
\non 
\ket{r,s,l} = \left(\begin{array}{c} 1 \\ 0  \end{array} \right) ,\quad  
\ket{r+\hbar,s,l} =\left( \begin{array}{c} 0\\1  \end{array} \right) ,
\ee 

\nin and used equation (\ref{eq:PxQx}) to perform a similar calculation. 

\nin
The eigenvalues of the matrix (\ref{eq:Ursl}) can be seen to be

\be
e^{i \omega_\pm}=\cos(\bar{L})\cos(\bar{K}) \pm i \sqrt{1-\cos^2(\bar{L})\cos^2(\bar{K})}.
\label{eq:eig_hbar_pi}
\ee

\nin
The surfaces $\omega_\pm (r,s)$  for $(K,L)=(\pi,\pi)$ and $(K,L)=(10,11)$ are plotted in Fig. \ref{fig:omega_pi}.  To emphasise that there may exist many different subspaces, labeled by $r$ and $s$, that give rise to eigenfunctions with same quasi-energy $\omega$ we have also illustrated some of the contour lines of $\omega_+$. 

The eigenvectors of (\ref{eq:Ursl}) may be written as 

\be
\label{eq:statpi}
\non 
\ket{\psi^\pm}=N_\pm \left( \begin{array}{c} \exp(-i\bar{L}) \sin(\bar{K})\theta \\ -\sin(\bar{L})\cos(\bar{K}) \mp \sqrt{1-\cos^2(\bar{L})\cos^2(\bar{K})} \end{array} \right) ,
\ee 

\nin where $N_{\pm}$ are normalizing coefficients. In the position representation these may be written as 

\be
\braket{q'}{\psi^\pm} =  \psi^\pm_{0,0} \braket{q'}{r,s,\pi} + \psi^\pm_{0,1} \braket{q'}{r,s+\hbar,\pi},
\label{eq:vecpi}
\ee

\begin{figure}[htbp]
   \centering
    \epsfig{file=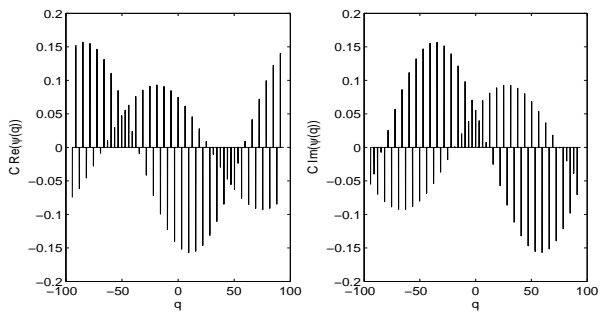,width=0.5\textwidth,height=0.25\textwidth }
\caption{An eigenstate viewed from the position representation, see (\ref{eq:vecpi}) and note \cite{C}. The stationary state is a superposition of two orthogonal $kq$ functions. }
   \label{fig:eigvecpi}
\end{figure}

\nin
where $\psi^\pm_{j,k}$ is given by $\braket{r +j\hbar,s+k\hbar,\pi}{\psi^\pm}$ and $\braket{q'}{r,s,l}$ is given by (\ref{eq:pos_rep}). An example of a single eigenvector, viewed from the position representation, is illustrated in Fig. \ref{fig:eigvecpi}.

\subsection{Numerical analysis with $\hbar = 2\pi/400$}

\begin{figure}[htbp]
   \centering
    \epsfig{file=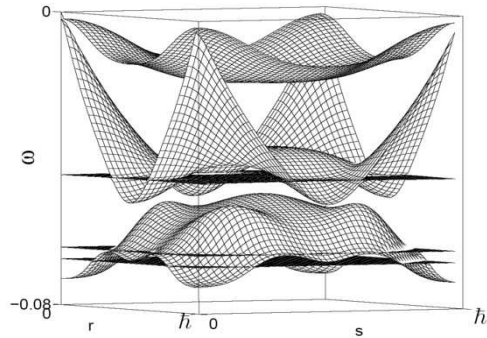,width=0.4\textwidth,height=0.25\textwidth }
\caption{Some quasi-energy surfaces with  $ K=2,L=2$ and $\hbar= 2 \pi/400$. The spectrum contains intersecting discrete and continuous bands.}
   \label{fig:hbar_pi_400}
\end{figure}

In a similar way we may calculate the matrix elements of (\ref{eq:xPxQ}) or (\ref{eq:PxQx}) for general $\hbar = 2 \pi M/N$. Explicit eigensolution calculations for large $N$ become unmanageable but can be done numerically using standard LAPACK routines.  As an example of this technique and the insight it provides into the system's quasi-energy structure we plot six quasi-energy bands as a function of $r$ and $s$, for $K=L=2$ and $\hbar= 2\pi/400$, in Fig. \ref{fig:hbar_pi_400}. The figure demonstrates that the quasi-energy structure of the system, under these parameters, contains intersecting point spectra {\em and} continuous bands.

\subsection{Convergents of $\hbar \ne 2\pi M/N$}

Block reduction can only take place for $\hbar = 2 \pi M/N$ as there is no way to choose $(l,a)$ so that they both are rational multiples of $2\pi$ when $\hbar =2\pi \rho$ and $\rho$ is irrational. However, it is possible to examine eigensolutions of the problem for rational approximates $M/N$ of $\rho$.  

\begin{figure}[htbp]
   \centering
    \epsfig{file=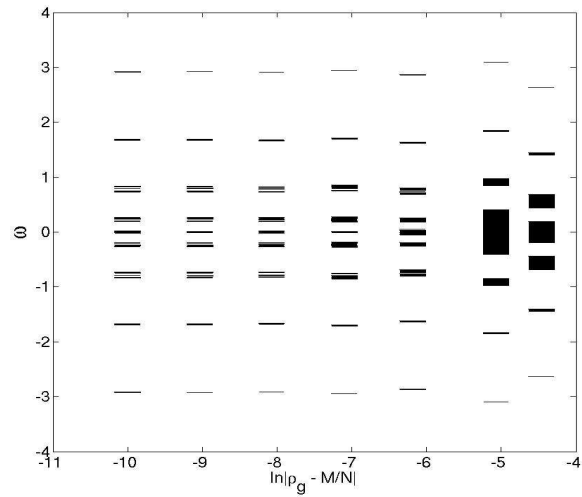,width=0.5\textwidth,height=0.4\textwidth }
\caption{The quasi-energy band structure  with $K=L=1.5$ and  $\hbar=2\pi M/N$ with $M/N = 1/7, 1/8, 2/15, 3/23, 5/38,8/61$ and $13/99$  These are the first 7 convergents of the continued fraction expansion of $\rho_g = 1/(6+\sigma_g)$ where $\sigma_g$ is the golden mean $(\sqrt{5}+1)/2$.} 
   \label{fig:eig_irat}
\end{figure}

%

 Following \cite{bor95b} we examine convergents of the continued fractions expansion of $ \hbar= 2\pi \rho_g$ where $\rho_g = 1/(6+\sigma_g)$ and $\sigma_g$ is the golden mean $(\sqrt{5}+1)/2$. The results are given graphically in Fig. \ref{fig:eig_irat}. The denominator $N$ of each convergent increases as we approach $\rho$ and therefore so must the number of quasi-energy bands. Recent analysis suggests that as $M/N \rightarrow \rho_g$ the spectral distribution may even become multi-fractal \cite{bor95a,ket99}.

\begin{figure}[htbp]
   \centering
    \epsfig{file=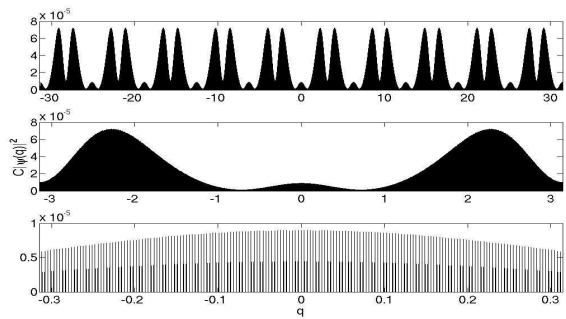,width=0.5\textwidth,height=0.25\textwidth }
\caption{An eigenfunction in the position basis with $K=1$ and $L=7$ with $\hbar=2 \pi \frac{233}{1775}$. To generate this state a superposition of the $N=1775$ $kq$ states $\braket{q'}{0,0+j\hbar,2\pi}$ with $j$ running from $0$ to $N$ was used, note \cite{C}.} 
   \label{fig:eigvec233_1775}
\end{figure}

As an example of a stationary state of a rational approximate to irrational $\rho_g$ we choose $M/N= 233/1775$. We plot one of the calculated eigenstates in the position representation in Fig. \ref{fig:eigvec233_1775}. At a glance the eigenfunction first appears to be continuous but on closer inspection it can be seen to be discrete. Importantly, this eigenstate and indeed all eigenstates calculated for a rational approximate to irrational $\rho$ are still periodic and still extended. There is no need to give them extra significance. 

\subsection{Quasi-probability distributions}

As already mentioned one may reconstruct each state in the position or momentum basis by using equations (\ref{eq:pos_rep}) or (\ref{eq:mom_rep}). However, we may also examine the Wigner and Husimi distributions of these states. As a brief example of this technique we plotted in Fig. \ref{fig:eigfunc_webq} the Husimi quasi-probability function of a stationary state that has the same web-structure as the classical Poincar\'{e} surface of section illustrated in Fig. \ref{fig:eigfunc_webcl}. The translational invariance of this stationary state is clearly visible although the Husimi distribution obscures it's discretised nature.

\begin{figure}
     
     \subfigure[Phase space plot of the accelerator modes responsible for anomalous diffusion with $K=L=2\pi+.1$]{
           \label{fig:eigfunc_acca} 
           \includegraphics[width=.30\textwidth,height=0.30\textwidth]{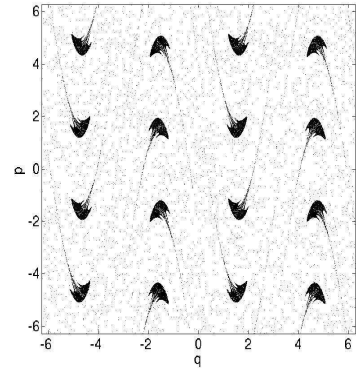}}
     \subfigure[The Husimi distribution of a stationary state for $K=L=2\pi+0.1$ and  $\hbar=2\pi/200$ ]{
           \label{fig:eigfunc_accb} 
           \includegraphics[width=.30\textwidth,height=0.30\textwidth]{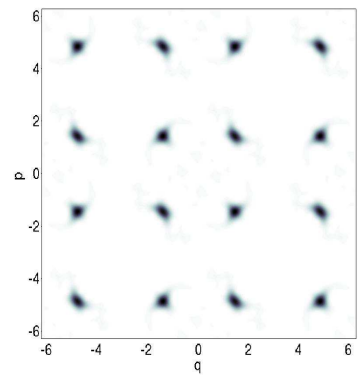}}

     \caption{Accelerator modes. The Husimi distribution of this stationary state has peaks primarily around the classical accelerator modes.  Note that Fig. \ref{fig:eigfunc_acca} is not a Poincar\'{e} surface of section but rather a plot of the initial conditions which, after $800$ iterations of the classical mapping, have energies $\frac{1}{2}(q^2 +p^2)> 3500$}
     \label{fig:eigfunc_acc}
\end{figure}

Now, we also briefly examine a stationary state responsible for quantum anomalous diffusion \cite{zas97}. Classical anomalous diffusion is due to stable phase space structures occurring for particular values of $K$ and $L$. When $K=L$ the stability conditions for the existence of these structures may be written as $2 n \pi < K < \sqrt{ (2n\pi)^2  +4}$ for integer $n$ \cite{kel05}. The plot in Fig. \ref{fig:eigfunc_acc} is evidence of a sort of scaring as classical points initially on an accelerator mode never return to the same region.

These examples illustrate how the quantum eigensolutions may resemble classical phase space structures. In further studies this may allow one to clearly distinguish between quasi-energy distributions associated with stable classical trajectories and those related to the unstable trajectories.

\section{CONCLUSION}

The Floquet operator for the kicked Harper model has been reduced to block matrices for $\hbar = 2 \pi N/M$ and positive integers $M$ and $N$. As each block is spanned by a set of $N$ orthogonal $kq$ $(rs)$ functions, any eigenfunction of the Floquet operator must be a superposition of this set and is therefore periodic in position and momentum.  

We have explicitly calculated the block matrix elements and showed that the matrices themselves may be written concisely using the generalised discrete Fourier transform. Using this method we have performed analysis on the system for various values of the quantum parameter $\hbar$ and the parameters $K$ and $L$. We have illustrated, for the first time, the 2-dimensional structure in the system's quasi-energy spectrum and demonstrated, albeit with just one example, that the distribution may be made up of intersecting discrete and continuous spectral bands. The behaviour of the spectrum as $\hbar$ approaches an irrational multiple of $2 \pi$ is also demonstrated. 

In addition to this, selected stationary states of the system have been calculated. It is demonstrated that they are, by construction, periodic and discrete in momentum and position. Two examples of stationary states whose Husimi distributions resemble classical phase space structures have also been provided. These results should help open up the way for a more detailed study of the system's eigensolutions and their relationship with the classical system.  
 
\section{Acknowledgments}
The author sincerely thanks D. M. Heffernan for the many fruitful discussions relating to this work.

\section{Appendix}

The elements of the Floquet matrix in the $kq$ representation may be written as

\bea
\bra{r'',s'',l} U \ket{r',s',l} =  && \sum_{r,s}\bra{r'',s'',l} P\ket{r,s,l} \\  &&\times  \bra{r,s,l} Q \ket{r',s',l}
\eea
 
\nin
We first calculate the expansion  

\bea
\label{eq:q_expand}
Q_{r,s,r',s'} &\equiv &\bra{r,s,l} Q \ket{r',s',l}   \\
&=& \int \int \braket{r,s,l}{q'} \bra{q'}Q(q) \ket{q''}\braket{q''}{r',s',l}dq' dq''. \non
\eea

\nin  By substituting (\ref{eq:pos_rep}) into (\ref{eq:q_expand}) and after some manipulation it can be seen that non-zero matrix elements occur only if $r=r'$. We therefore write   

\bea
\label{eq:Qrs1}
Q_{r,s,r,s'} &=& C_q^2 \sum_{n} \exp \left(-\frac{is}{\hbar}(\frac{r}{2}+nl)\right) \times \\ && \sum_m Q(r+ml) \exp \left(\frac{is'}{\hbar}(\frac{r}{2}+ml)   \right). \non
\eea  

\nin
It is now that the choice of $(l,a)$ becomes important for the block reduction. Setting $(l,a) = (2\pi M /N_s, 2 \pi/N_r)$ one can see after some consideration, that the summations are only well-defined if  $s-s'=(j-k)\hbar$ with $j,k=0...N_s-1$. To this end we make the substitutions $s=s+k\hbar$ and $s'=s+j\hbar$ to get

\bea
\label{eq:Qrs2}
Q_{r,s} &=& C_q^2\sum_{n} \exp \left(- i (k + \frac{s}{\hbar})(\frac{r}{2l}+n)l\right) \times \\ && \sum_m Q(r+ml) \exp \left(i (j + \frac{s}{\hbar})(\frac{r}{2l}+m)l  \right). \non
\eea  

\nin A simple analysis shows that the two infinite summations can be reduced to finite ones with integers $n$ and $ m$ running from $0$ to $N_s-1$. By representing the orthonormal $kq$ kets as

\bea  
&& \ket{r,s,l} \quad \quad \quad= (1,0,0,0, ... ,0,0)^T \non \\
&& \ket{r,s+\hbar,l}\;\;\;\;\;= (0,1,0,0, ... ,0,0)^T \non \\
&& \ket{r,s+2\hbar,l}\;\;\;\;\;= (0,0,1,0, ...,0,0)^T \non \\
&& \quad \quad  \quad \quad\quad \quad  \quad \vdots \non \\
&& \ket{r,s+(N_s -1)\hbar,l}= (0,0,0,0, ...,0,1)^T ,\non 
\eea

\nin
one can write these finite summations as matrix multiplications. That is  
 
\be
Q(r,s) = C_q^2 F \left(\frac{s}{\hbar}, \frac{r}{2l}\right) D(V)  \; F \left(\frac{s}{\hbar}, \frac{r}{2l}\right)^\dag \;\;
\label{eq:Q}
\ee

\nin
where $D$ is the diagonal matrix defined by

\be
\mathcal{D}(V)_{j,j}= \exp \left(-\frac{i}{\hbar} V(r+jl) \right),   
\label{eq:UQ}
\ee

\nin
and $F(a,b)$ is the generalised Fourier matrix 

\be
F_{j,k} = \omega_q^{(j+a)(k+b)}, 
\ee

\nin
 with $\omega_q=\exp(-i 2\pi M /N_s)/\sqrt{N_s}$. 

A similar analysis can be done for the matrix elements $\bra{r'',s'',l}P\ket{r,s,l}$ by expanding in the momentum basis giving

\bea
P(r,s)= C_p^2 F(\frac{s}{2a},\frac{r}{\hbar})^\dag D(W) \; F(\frac{s}{2a},\frac{r}{\hbar})
\label{eq:P}
\eea

 \nin
where in this case the matrix $D$ is defined as 

\be
D(W)_{j,j}= \exp \left(-\frac{i}{\hbar} W(s+ja) \right),  
\ee

\nin
 and $\omega_p= \exp(-i 2\pi/N_r)/ \sqrt{N_r}$.  Of course to write it like this the kets spanning the subspaces are represented as

\bea  
&& \ket{r,s,l} \quad \quad \quad= (1,0,0,0, ... ,0,0)^T \non \\
&& \ket{r+\hbar,s,l}\;\;\;\;\;= (0,1,0,0, ... ,0,0)^T \non \\
&& \ket{r+2\hbar,s,l}\;\;\;\;\;= (0,0,1,0, ...,0,0)^T \non \\
&& \quad \quad  \quad \quad\quad \quad  \quad \vdots \non \\
&& \ket{r+ (N_r-1)\hbar,s,l}= (0,0,0,0, ...,0,1)^T \non .
\eea

\nin
It should be noted here that the expression for $\omega_p$ does not contain the factor $M$. However, this is only a consequence of our choice $(l,a)$. If one sets  $(l,a) = (2\pi/N_s, 2 \pi M /N_r)$ we get $\omega_q=\exp(-i 2\pi /N_s)/\sqrt{N_s}$ and $\omega_p=\exp(-i 2\pi M /N_r)/\sqrt{N_r}$.

\vspace{3mm}
Together $Q(r,s)$ and $P(r,s)$ act only on the subspaces spanned by the kets $\ket{r+j\hbar,s+k\hbar,l}$ for $j=0...N_r -1$ and $k=0....N_s -1$. We define the the block diagonal matrix $\mathcal{Q}$ as

\bea
\mathcal{Q}(r,s) &\equiv& \left[ \begin{array}{cccc} Q(r_1,s) & 0 & \cdots & 0\\  0 & Q(r_2,s) & \cdots & 0 \\  \vdots & \vdots & \ddots & \vdots\\ 0 & 0 & \cdots & Q(r_{N_r},s) \end{array} \right],
\label{eq:Qmat}
\eea

\nin where $r_n=r+(n-1)\hbar$ and each block matrix $Q(r_n,s)$ is given by (\ref{eq:Q}). This matrix calculates the operation $Q(q)=e^{\frac{-i}{\hbar} V(q)} $ over the whole subspace. In the same way we define the block diagonal matrix $\mathcal{P}$ as 
\bea
\mathcal{P}(r,s) &=& \left[ \begin{array}{cccc} P(r,s_1) & 0 & \cdots & 0\\  0 & P(r,s_2) & \cdots & 0 \\  \vdots & \vdots & \ddots & \vdots\\ 0 & 0 & \cdots & P(r,s_{N_s}) \end{array} \right], 
\label{eq:Pmat}
\eea

\nin for the $e^{\frac{-i}{\hbar}W(p)} $ operation. Here, $s_n=s+(n-1)\hbar$ and each block matrix $P(r,s_n)$ is given by (\ref{eq:P}). It is helpful to note here that $C_q^2 C_p^2 =1$.

In order to multiply these block matrices together one must first re-order the elements of one of the matrices. This can be accomplished through the use of a simple swap matrix. If $n$ is the $n^{th}$ column vector of the $N \times N$ identity matrix we can define the swap matrix $X(j,k)$ as

\bea
X(j,k)=[1,1+j,&&... \;1+(k-1)j, \;2,\;2+j, \; ...\non  \\ 
&&... \;j+(k-2)j, j\times  k] ,
\eea

\nin where of course $j\times k=N$. If we retain the basis labeling of (\ref{eq:Qmat}) we must rearrange $P(s,r)$ to get

\be
\mathcal{P}'(r,s)= X(N_r,N_s)^\dag \mathcal{P}(r,s) X(N_r,N_s) .
\ee

\nin
The total evolution operator, (\ref{eq:U_KHM}), on the $r,s$ subspace, can then be written as: 

\be
U(r,s) = \mathcal{P}'(r,s) \mathcal{Q}(r,s) \;.
\label{eq:xPxQ}
\ee

\nin
However, we could just as easily have retained the basis labeling of (\ref{eq:Pmat}) by reordering  $\mathcal{Q}(r,s)$ to get

\be
\mathcal{Q}'(r,s)= X(N_r,N_s) \mathcal{Q}(r,s) X(N_r,N_s)^{\dag}, 
\ee  

\nin
and now the Floquet operator is written as

\be
U(r,s) = \mathcal{P}(r,s)  \mathcal{Q}'(r,s) .
\label{eq:PxQx}
\ee

It is the special case of when $N_s=1$ and $N_r=N$ that gives us method used in \cite{ket99,sat05} to analyse this system.  The matrix $\mathcal{Q}(r,s)$ reduces to one with just $\exp(-i V(r_n)/\hbar)$ along the diagonal and  $\mathcal{P}(r,s)$ is given by (\ref{eq:P}). In total we have

\bea \non
U(r,s) &=& \mathcal{P}(r,s) \mathcal{Q}'(r,s) = \mathcal{P}(r,s) \mathcal{Q}(r,s) \;\;,\\  &=& F(\frac{s}{2a},\frac{r}{\hbar})^\dag  \exp \left(-\frac{i}{\hbar} W(s_n) \right) \\  \non& & \;\;  \times \;\; F(\frac{s}{2a},\frac{r}{\hbar}) \exp \left(-\frac{i}{\hbar} V(r_n) \right). 
\label{eq:Sat}
\eea

\nin Equation (\ref{eq:Sat}) bears a striking similarity to the position representation of $U$ in (\ref{eq:U}), albeit with discrete values of $q$. Understandably, this often led to the interpretation that the eigenvector elements of $U(r,s)$ are discrete values of the {\em actual} eigenfunctions of $U$ in the position basis. However, as this analysis has just shown, each block matrix is spanned by the vectors 
\be
\ket{r+j\hbar,s+k\hbar,l}
\label{eq:span2}
\ee

\nin  where $j \in {0,1,...,N_r-1}$ and  $k \in {0,1,...,N_s-1}$. Therefore solutions to the eigenvalue equation, $U\ket{\psi} = e^{i\omega} \ket{\psi}$, must be of the form

\be
\ket{\psi} = \sum_{j =0}^{N_r -1} \sum_{k=0}^{N_s-1} \psi_{j,k} \ket{r+j\hbar,s+k\hbar,l}.
\label{eq:eig2}
\ee

An eigenstate viewed from the position representation $\braket{q'}{\psi}$ is therefore expanded as
 
\be
\braket{q'}{\psi} = \sum_{j =0}^{N_r-1} \sum_{k=0}^{N_s-1} \psi_{j,k} \braket{q'}{r+j\hbar,s+k\hbar,l}.
\ee

\nin where $\psi_{j,k}=  \braket{r+ j\hbar,s+k\hbar,l }{\psi}$ and $\braket{q'}{r+j\hbar,s+k\hbar,l}$ is defined by(\ref{eq:pos_rep}). By construction the values $\psi_{j,k}$ are the individual elements of the eigenvectors of the matrices $U(r,s)$.


\begin{thebibliography}{99}
\bibitem{poisson} We have used the Poisson sum formula $\sum_{n=-\infty}^{\infty}  \exp \left( -\frac{i 2\pi p n }{a} \right) = a \sum_{m=-\infty}^{\infty} \delta(p - m a)$ to carry  out this calculation.
\bibitem{C} The constant $C$ is introduced so that, using (\ref{eq:U}) and the Fast Fourier transform, we can numerically check that the constructed functions are eigenfunctions of $U$ in the position representation. The value of $C$ is unimportant and only depends on how much of the vector one wants to examine.
\end{thebibliography}
\end{document}